\documentclass[aps,prb,twocolumn,showpacs,superscriptaddress]{revtex4-1}

\usepackage{amsmath}
\usepackage{amssymb}
\usepackage{bm}
\usepackage{textcomp}
\usepackage{graphicx}
\usepackage{bbold}
\usepackage[english]{babel}
\usepackage[usenames,dvipsnames]{xcolor}
\usepackage{gensymb} %for degree symbol

\newcommand{\bea}{\begin{eqnarray*}}
\newcommand{\eea}{\end{eqnarray*}}
\newcommand{\bne}{\begin{equation*}}
\newcommand{\ede}{\end{equation*}}

\newcommand{\bnen}{\begin{equation}}
\newcommand{\eden}{\end{equation}}
\newcommand{\bean}{\begin{eqnarray}}
\newcommand{\eean}{\end{eqnarray}}
\newcommand{\bnsn}{\begin{subequations}}
\newcommand{\edsn}{\end{subequations}}

\newcommand{\bna}{\begin{array}}
\newcommand{\eda}{\end{array}}
\newcommand{\bnm}{\begin{enumerate}}
\newcommand{\edm}{\end{enumerate}}
\newcommand{\bni}{\begin{itemize}}
\newcommand{\edi}{\end{itemize}}

\renewcommand{\vec}[1]{\text{\boldmath{$ #1 $}}}

\begin{document}

\title{Spin-strain interaction in
nitrogen-vacancy centers in diamond}

\author{P\'eter Udvarhelyi}
\affiliation{E\"otv\"os University, H-1117 Budapest, Hungary}
\affiliation{Wigner Research center for Physics, Hungarian Academy of Sciences, 
PO. Box 49, H-1525, Budapest, Hungary}
\author{Vladyslav O. Shkolnikov}
\affiliation{Department of Physics, University of Konstanz, D-78457 Konstanz, Germany}
%\author{Adrian Auer}
%\email{adrian.auer@uni-konstanz.de}
\author{Adam Gali}
\affiliation{Wigner Research center for Physics, Hungarian Academy of Sciences, 
PO. Box 49, H-1525, Budapest, Hungary}
\affiliation{Department of Atomic Physics, Budapest University of Technology and Economics, Budafoki \'ut 8., H-1111 Budapest, Hungary}
\author{Guido Burkard}
\affiliation{Department of Physics, University of Konstanz, D-78457 Konstanz, Germany}
\author{Andr\'as P\'alyi}
\affiliation{Department of Physics, Budapest University of Technology and Economics, Budafoki \'ut 8., H-1111 Budapest, Hungary}
\affiliation{MTA-BME Exotic Quantum Phases "Momentum" Research Group, Budapest University of Technology and Economics, Budafoki \'ut 8., H-1111 Budapest, Hungary}

\begin{abstract}
The interaction of solid-state electronic spins with deformations
of their host crystal is an important ingredient in many
experiments realizing quantum information processing schemes.
Here, we theoretically characterize that interaction for a 
nitrogen-vacancy (NV) center in diamond. 
We derive the symmetry-allowed Hamiltonian describing 
the interaction between the ground-state spin-triplet
electronic configuration and the local strain. 
We numerically calculate the
six coupling-strength parameters of the Hamiltonian 
using density  functional theory, 
and propose an experimental setup
for measuring those coupling strengths. 
The importance of this interaction is highlighted by the fact
that it enables 
to drive spin transitions, 
both magnetically allowed and forbidden, 
via
mechanically or electrically driven spin resonance.
This means that the ac magnetic field routinely used in a wide range of spin-resonance experiments with NV centers could in principle be replaced by ac strain or ac electric field, potentially offering lower power requirements, simplified device layouts, faster spin control, and local addressability of electronic spin qubits.
\end{abstract}

%\pacs{}

\maketitle

\newcommand{\bra}[1]{\langle #1|}
\newcommand{\ket}[1]{|#1\rangle}
\newcommand{\braket}[2]{\langle #1|#2\rangle}
\newcommand{\nonum}{\nonumber \\}
\newcommand{\C}[1]{${}^{\textrm{#1}}$C}
\newcommand{\N}[1]{${}^{\textrm{#1}}$N}
\newcommand{\noteandras}[1]{\textcolor{blue}{#1}}

%\tableofcontents

%\textcolor{red}{
%Dear All:
%Please read the manuscript. 
%I tried to make this version such that
%it is submittable even if section V, i.e., the 
%"Schemes to measure the
%spin-strain parameters"
%section, is not completed.
%This might be useful if we have to rush the submission
%due to competition. }

%\textcolor{red}{
%Please let me know if you're okay with the completed parts
%(i.e., everything but section V.), or if you have suggestions
%you'd like me to implement (corrections, new figs, etc), 
%or doubts you'd like to discuss,
%or you'd like to add and/or rewrite parts, 
%e.g., to make the content stronger, etc.}

%\textcolor{red}{
%I think the next steps are to 
%(1) finalize the completed parts as suggested above,
%(2) work further toward results to be presented in section V.
%}

%\textcolor{red}{
%Looking forward to your feedback. 
%}

\section{Introduction}

\begin{figure}
\includegraphics[width=0.7\columnwidth]{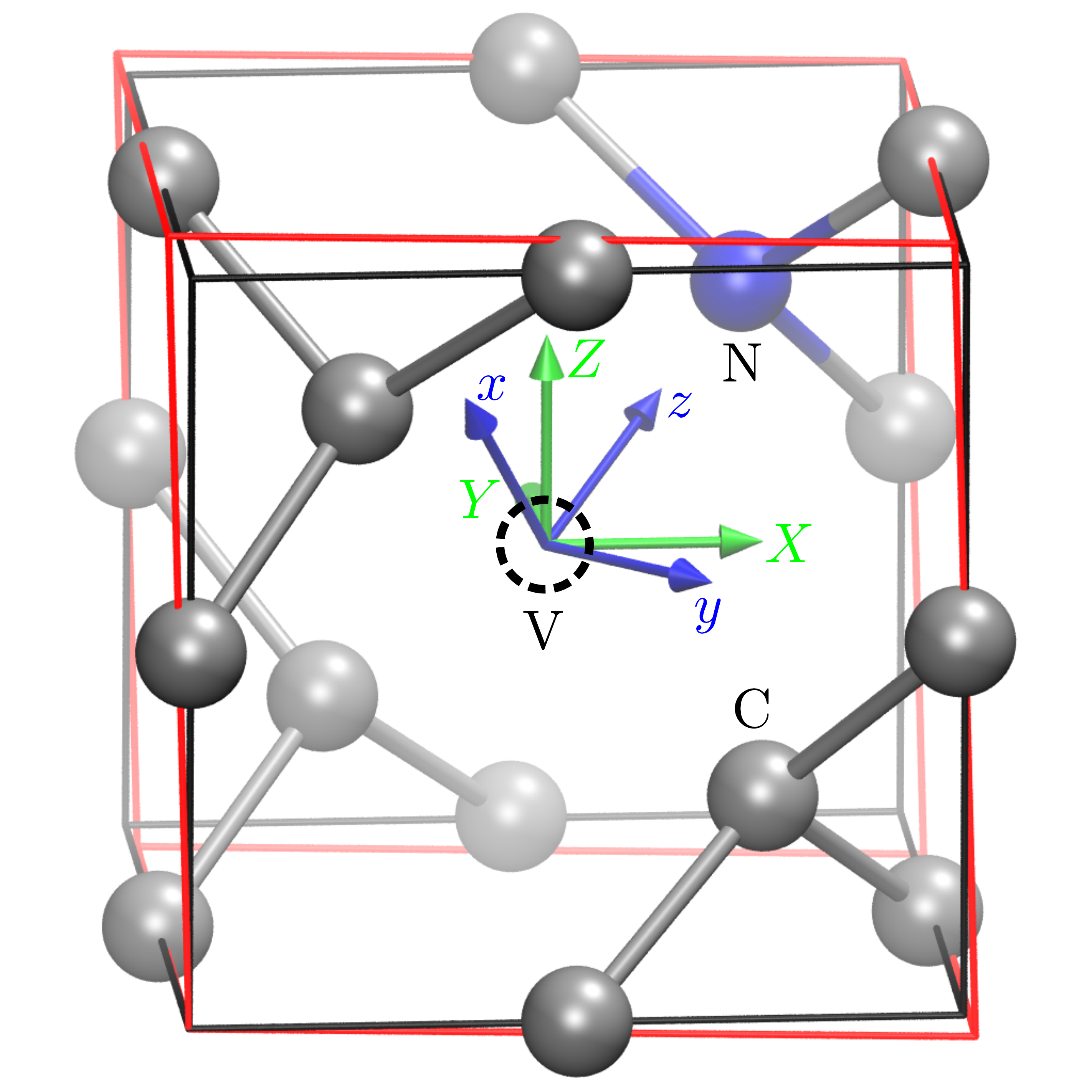}
\caption{\label{fig:NV}Nitrogen-vacancy (NV) center in the
  diamond lattice (Bravais cell depicted as a cube in
  black). $\{X,Y,Z\}$ defines the cubic reference frame and $\{x,y,z\}$
  defines the NV reference frame. Deformation of the diamond crystal is visualized in red for $\varepsilon_{xx}=0.1$ strain component. We use this high strain only for sake of clarity.}
\end{figure}

The nitrogen-vacancy (NV) color center
consists of a nitrogen atom substituting a carbon atom adjacent to a vacancy  
in diamond (see Fig.~\ref{fig:NV}). In the negatively charged state,
it shows a broad fluorescence with zero-phonon-line at
637~nm~\cite{duPreez:1965, Davies1976} and possesses a spin $S=1$
ground state~\cite{Loubser1977, Goss1996, Manson2006, Gali2008}. The electron spin of the NV center can be
initialized, coherently manipulated, and read out
in optically
detected magnetic resonance (ODMR) experiments\cite{Jelezko2004}, 
even at the level of individual centers\cite{Gruber1997}. 
This electronic spin degree of freedom is robust even at
room temperature, and its coherence time is typically a
few microseconds in natural
diamond~\cite{Jelezko2004}, reaching milliseconds in $^{12}$C
enriched diamonds~\cite{Balasubramanian:NatMat2009}. Because of these
favorable properties of the NV center, it provides a versatile
and highly coherent  
platform for the experimental realization of 
many quantum information schemes.
To maximize the potential of these defects
for various quantum communication~\cite{Bernien2013, Hensen2015, Kalb2017}, quantum sensing~\cite{Balasubramanian2008, Maze2008, Dolde2011, Toyli2013, Neumann2013, Kucsko2013, TeissierPRL, Barfuss:2015, Ovartchaiyapong_2014},
and quantum computing~\cite{Cramer,Waldherr,Taminiau} 
applications,
it is crucial to understand the interaction of the center's electronic
system with its environment, most notably externally induced
electromagnetic fields, and deformations of the crystal lattice.

In this work, we provide a theoretical description of the latter,
i.e., the 
\emph{spin-strain interaction Hamiltonian} of the  
spin-triplet ground-state electronic configuration
of the negatively charged NV defect.
Even though in recent years this interaction has been studied 
intensively\cite{MazeNJP,DohertyNJP,DohertyPRB2012,Doherty20131,Cai,TeissierPRL,Barfuss:2015,MacQuarrie:15,Ovartchaiyapong_2014,MacQuarriePRL, MacQuarrie_NatComm2017,MacQuarrie_dynamical,BarsonNL,Golter_PRL2016,GolterPRX,Meesala}, 
to our knowledge the correct and complete
form of the interaction Hamiltonian of the 
ground-state spin-1 vector $\vec S$ and the $3\times 3$
strain tensor $\varepsilon$
has not been established in the literature.
To fill this gap, we first derive the symmetry-allowed 
form of the spin-strain 
interaction Hamiltonian, see 
Eq.~\eqref{eq:spinstrain}.
Second, we present numerical results for the six coupling-strength 
parameters appearing in the spin-strain interaction Hamiltonian
(see Table \ref{tab:DFTresults}),
which we compute using density functional theory (DFT);
we find reasonable agreement with experimental results
(see Table \ref{tab:comparison}). 
Third, we propose a setup to measure
those two coupling-strength parameters which have not 
been experimentally characterized yet
(Section \ref{sec:exp}).
Finally, we discuss how the spin-strain interaction 
can contribute to various applications of NVs in 
quantum information schemes
(Section \ref{sec:discussion}).
In particular, 
our results reveal the possibility of using electric signals 
to control the magnetically allowed spin transitions of these
defects, potentially offering lower power requirements, 
simplified device layouts, faster spin control, and
local addressability of spin qubits.

We formulate our results in terms of the 
spin-stress interaction as well.
We emphasize that our qualitative considerations
apply more generally, to the whole family of 
spin-1 electronic states of defects with
$C_{3v}$ symmetry.

\section{Preliminaries}
\label{sec:preliminaries}

We choose the \emph{cubic reference frame}
such that its origin coincides with the vacancy, 
and the nitrogen is at $(a/4,a/4,a/4)$, with
$a$ being the width of the cubic cell.
The coordinates in the cubic frame are referred to as
$X$, $Y$, $Z$ (see Fig.~\ref{fig:NV}). 
The \emph{NV reference frame} is defined via its three 
orthonormal basis vectors, 
$\vec e_z = (1,1,1)/\sqrt{3}$,
$\vec e_y = (1,-1,0)/\sqrt{2}$,
and $\vec e_x = \vec e_y \times \vec e_z$. 
From now on, unless noted otherwise, we use the 
NV frame, and $x$, $y$, and $z$ refers to coordinates
in the NV frame.
This choice of the reference frame 
implies that reflection upon the $xz$ plane is
a symmetry of the structure. 
This reflection, together with the 3-fold rotation around the $z$ axis, 
generate the point group $C_{3v}$ of the defect. 

In the presence of a homogeneous magnetic field 
$\vec B = (B_x,B_y,B_z)$,
and in the absence of any electric field and strain, 
the NV spin is described by the following Hamiltonian:
\bean
\label{eq:he}
H_e/h = D S_z^2 + \gamma_e \vec B \cdot \vec S,
\eean
where 
$h$ is Planck's constant, 
$D = 2.87 \, \text{GHz}$ is the zero-field splitting,
$\gamma_e= 2.8 \, \text{MHz}/\text{G}$ 
is the electron gyromagnetic ratio,
and 
$\vec S = (S_x,S_y,S_z)$ is the vector of
spin-1 Pauli matrices.
The eigenstates and eigenvalues of $S_z$ will 
be labelled according to $S_z \ket{m_S e} = m_S \ket{m_S e}$,
where $m_S \in \{-1,0,+1\}$.

The interaction Hamiltonian of a homogeneous electric field 
$\vec E=(E_x,E_y,E_z)$ 
with the NV spin is constrained by the
$C_{3v}$ symmetry of the defect, and hence described 
by\cite{vanOort,DohertyPRB2012,Matsumoto} 
\begin{subequations}
\label{eq:electrichamiltonian}
\bean
H_E &=& H_{E0} + H_{E1} + H_{E2}, \\
H_{E0}/h &=& d_\parallel S_z^2 E_z,\\
H_{E1}/h&=& d'_\perp \left[
  \{S_x,S_z\} E_x + 
    \{S_y,S_z\} E_y
\right], \\
H_{E2}/h &=& d_\perp \left[
  (S_y^2-S_x^2) E_x
  +\{S_x,S_y\} E_y
\right].
\eean
\end{subequations}
Here, the lower indices 0, 1 and 2 refer to the difference
in the electron spin quantum numbers ($m_S$)
connected by the corresponding Hamiltonian; 
e.g., $H_{E1}$ has nonzero matrix elements between
$\ket{0e}$ and $\ket{\pm1e}$.
The coefficients 
$d_\perp = 17 \, \text{Hz}\, \text{cm}/\text{V}$ and 
$d_\parallel = 0.35\, \text{Hz}\, \text{cm}/\text{V}$ 
have been inferred
in the experiment of Ref.~\onlinecite{vanOort}.
However, to our knowledge, the coefficient $d'_\perp$ has not
been quantified experimentally or theoretically;
nevertheless it is expected\cite{DohertyPRB2012}
to have the same order of magnitude as $d_\perp$. 

Two remarks on the spin-electric interaction Hamiltonian $H_E$:
(1) The presence of $H_{E1}$ in the spin-electric Hamiltonian
is a clear indication that coherent Rabi oscillations within
the state pairs
$\ket{0e} \leftrightarrow \ket{+1e}$
and 
$\ket{0e} \leftrightarrow \ket{-1e}$ 
can be driven by an ac electric field.
This means, in principle, that any coherent-control experiment where
these transitions are driven by ac magnetic field
can also be done by replacing the ac magnetic field 
with an ac electric field, e.g., created by a single metallic gate electrode.
To our knowledge, this opportunity which is routinely exploited
for various solid-state spin systems\cite{Kato,Nowack,Golovach,Klimov} 
and is known as
\emph{electrically driven spin resonance} 
or
\emph{electric dipole spin resonance},
has been overlooked
in the literature in the context of the magnetically
allowed 
$\ket{0e} \leftrightarrow \ket{+1e}$
and
$\ket{0e} \leftrightarrow \ket{-1}$
transitions of NVs and similar 
defects with $C_{3v}$ symmetry. 
Since electric control might bring significant advantages 
over magnetic control
(simplified device layout, well-confined control fields allowing
for local spin addressability, lower power requirements, etc),
this observation provides a strong motivation to 
characterize the coupling-strength parameter $d'_\perp$ of
$H_{E1}$ both experimentally and theoretically.
(2) The experimental setup we propose in 
Sec.~\ref{sec:exp} to measure spin-stress and
spin-strain   coupling-strength
parameters can be easily adopted to measure $d'_\perp$.

\section{Spin-strain Hamiltonian}
\label{sec:symmetry}

In our understanding, the spin-strain interaction Hamiltonians
used in the literature to characterize the NV
(and similar defects with $C_{3v}$ symmetry)
are incomplete.
A central result of the present work is the 
most general 
form of this Hamiltonian that is compatible with the $C_{3v}$
symmetry
of the NV. 
We find that this general symmetry-allowed Hamiltonian is 
characterized by six independent real 
\emph{coupling-strength parameters}
$h_{41}$, $h_{43}$, $h_{25}$, $h_{26}$, $h_{15}$, $h_{16}$,
and 
has the following form: 
\begin{subequations}
\label{eq:spinstrain}
\bean
H_{\varepsilon} &=& H_{\varepsilon 0} + H_{\varepsilon 1}+H_{\varepsilon 2},
\\
H_{\varepsilon 0} /h&=&
[h_{41} (\varepsilon_{xx} + \varepsilon_{yy}) 
+
h_{43} \varepsilon_{zz}]S_z^2,
\label{eq:h0}
\\
\label{eq:h1}
H_{ \varepsilon 1}/h &=& \nonumber
\frac 1 2
\left[h_{26} \varepsilon_{zx}
- \frac 1 2 h_{25} (\varepsilon_{xx} - \varepsilon_{yy})
\right]
 \{ S_x,S_z\}
\\
&+&
\frac 1 2 
\left(
h_{26}  \varepsilon_{yz} 
+  h_{25} \varepsilon_{xy}
\right)
\{ S_y,S_z\},
\\
H_{\varepsilon 2} /h&=&  \frac 1 2 \left[
	 h_{16}  \varepsilon_{zx}
	- \frac 1 2 h_{15} (\varepsilon_{xx} - \varepsilon_{yy})
\right](S_y^2-S_x^2) \nonumber
\\
&+& \frac 1 2 (
 h_{16} \varepsilon_{yz} + h_{15} \varepsilon_{xy}
) \{S_x, S_y\},
\eean
\end{subequations}
where $\varepsilon_{ij} = (\partial u_i/\partial x_j +\partial
u_j/\partial x_i)/2$ denotes the strain tensor and ${\bf u}({\bf r})$
is 
the displacement field.
Similarly to Eq.~\eqref{eq:electrichamiltonian},
the subscripts 0, 1, and 2 here refer to the
difference in the electron spin quantum numbers $m_S$ connected
by the corresponding Hamiltonian.
We present an elementary derivation of Eq.~\eqref{eq:spinstrain},
as well as a derivation based on group representation theory,
in Appendix \ref{app:symmetry}.

Note that the symmetry-allowed form of the spin-stress interaction, 
i.e., when the mechanical deformation is characterized by 
the $3\times 3$ stress tensor $\sigma$ instead of strain $\varepsilon$, 
is completely analogous to Eq.~\eqref{eq:spinstrain}.
In what follows, we adopt a notation for the spin-stress
Hamiltonian $H_\sigma$ that is  analogous to 
Eq.~\eqref{eq:spinstrain}, with the 
substitutions $\varepsilon \mapsto \sigma$ 
and $h \mapsto g$:
\begin{subequations}
\begin{eqnarray}
\label{eq:GSLAC_H}
%\begin{split}
H_{\sigma}&=& H_{\sigma0} + H_{\sigma 1} + H_{\sigma 2}, \\
H_{\sigma 0}/h&=&\left[g_{41}(\sigma_{xx}+\sigma_{yy})+g_{43}\sigma_{zz}\right]S_z^2,
\\
H_{\sigma 1}/h&=&\frac{1}{2}\left[g_{26}\sigma_{xz}-\frac{1}{2}g_{25}(\sigma_{xx}-\sigma_{yy})\right]\{S_x,S_z\}
\nonumber
\\&+&\frac{1}{2}(g_{26}\sigma_{yz}+g_{25}\sigma_{xy})\{S_y,S_z\},
\\
H_{\sigma 2}/h&=&\frac{1}{2}\left[g_{16}\sigma_{xz}-\frac{1}{2}g_{15}(\sigma_{xx}-\sigma_{yy})\right](S_y^2-S_x^2)
\nonumber
\\&+&\frac{1}{2}(g_{16}\sigma_{yz}+g_{15}\sigma_{xy})\{S_x,S_y\}.
\end{eqnarray}
\label{Stress_ham}
\end{subequations}

Many recent works
(e.g., Refs. \onlinecite{TeissierPRL,Ovartchaiyapong_2014,MacQuarriePRL})
rely on a heuristic spin-strain Hamiltonians 
built on an unjustified analogy between strain and 
electric field. 
That approach does not take into account the
$3\times 3$ tensor
structure of strain, therefore it provides an incorrect description
of the spin-strain interaction, even in the absence shear strain. 
A recent work\cite{BarsonNL} 
uses a spin-stress Hamiltonian based on the 
$3\times 3$ stress tensor $\sigma$; 
their Hamiltonian includes 4 real parameters, $a_1$, 
$a_2$, $b$ and $c$. 
That Hamiltonian is equivalent
to our $H_{\sigma 0} + H_{\sigma 2}$;
but incomplete as it lacks
the symmetry-allowed term $H_{\sigma 1}$ analogous to 
Eq.~\eqref{eq:h1};
we provide more details on 
its relation to our results
in Sec.~\ref{sec:dft}.
We note that using the incomplete $H_{\sigma 0}+H_{\sigma 2}$
Hamiltonian in Ref.~\onlinecite{BarsonNL} is justified as an 
approximation, since the
term $H_{\sigma1}$ is a small perturbation in the magnetic-field
range addressed in those experiments.
We also remark that in a very recent work\cite{Norambuena}, 
a spin-phonon interaction Hamiltonian incorporating
matrix elements between $\ket{0e}$ and $\ket{\pm 1 e}$
has been used to describe spin relaxation in NVs.

%====

\section{Spin-strain parameters from 
density functional theory}
\label{sec:dft}

\begin{table}
\caption{
 Spin-strain ($h$) 
and spin-stress ($g$) coupling-strength parameters
calculated from density functional theory.
See Appendix \ref{app:dft} for methodological details. 
Results are rounded to significant digits.}
\begin{ruledtabular}
\begin{tabular}{c|c|c|c}
parameter & value (MHz/strain) & parameter & value (MHz/GPa)\\\hline
$h_{43}$ & $2300\pm200$ & $g_{43}$ & $2.4\pm0.2$\\
$h_{41}$ & $-6420\pm90$ & $g_{41}$ & $-5.17\pm0.07$\\
$h_{25}$ & $-2600\pm80$ & $g_{25}$ & $-2.17\pm0.07$\\
$h_{26}$ & $-2830\pm70$ & $g_{26}$ & $-2.58\pm0.06$\\
$h_{15}$ & $5700\pm200$ & $g_{15}$ & $3.6\pm0.1$\\
$h_{16}$ & $19660\pm90$ & $g_{16}$ & $18.98\pm0.09$\\
\end{tabular}
\end{ruledtabular}
\label{tab:DFTresults}
\end{table}

We use DFT to numerically compute the six coupling-strength coefficients
$h_{41}$, etc., appearing in the spin-strain Hamiltonian 
\eqref{eq:spinstrain}.
Methodological details are presented in Appendix \ref{app:dft}.
The results are summarized in Table \ref{tab:DFTresults}.
Therein, we also present
the spin-stress coupling-strength
coefficients $g_{41}$, etc, which we obtain from
the $h$ values using the stiffness tensor 
of bulk diamond, see Appendix \ref{app:conversion}.

In Table \ref{tab:comparison}, we compare the numerical DFT 
results of Table \ref{tab:DFTresults}
to the experimental results of Ref.~\onlinecite{BarsonNL}.
In Ref.~\onlinecite{BarsonNL},
four out of the six independent spin-stress coupling-strength
parameters of the spin-stress interaction Hamiltonian
were measured. 
Ref.~\onlinecite{BarsonNL} 
defines these 4 spin-stress coupling-strength parameters,
denoted as $a_1$, $a_2$, $b$, $c$, 
in a `hybrid' representation, where the spin-stress
Hamiltonian is expressed in terms of the NV-frame components
of the spin vector ($S_x$, $S_y$, $S_z$) and 
the cubic-frame components
of the stress tensor ($\sigma_{XX}$, $\sigma_{XY}$, etc).
To be able to make a comparison between our DFT results
and the experimental ones, 
we now take the notations of Ref.~\onlinecite{BarsonNL},
and introduce $d$, $e$, $\mathcal{N}_x$,
$\mathcal{N}_y$, to express
our spin-stress Hamiltonian 
$H_{\sigma}$ in Eq.~\eqref{Stress_ham}
in this hybrid representation:
\begin{subequations}
\bean
H_{\sigma 0} /h &=& \mathcal M_z S_z^2, \\
H_{\sigma 1} /h &=& 
  \mathcal{N}_x \{S_x,S_z\} 
  + \mathcal{N}_y \{S_y,S_z\},
\\
H_{\sigma 2} /h &=& 
  - \mathcal{M}_x (S_x^2-S_y^2)
  + \mathcal{M}_y \{S_x,S_y\}, 
\eean
\end{subequations}
where
\begin{subequations}
\bean
\mathcal{M}_z &=&
a_1 (\sigma_{XX} + \sigma_{YY} + \sigma_{ZZ}) \nonumber \\
&+& 
2a_2 (\sigma_{YZ} + \sigma_{ZX} + \sigma_{XY}),
\\
\mathcal{N}_x &=&
  d(2 \sigma_{ZZ} - \sigma_{XX} - \sigma_{YY}) \nonumber \\
  &+&e(2\sigma_{XY} - \sigma_{YZ} - \sigma_{ZX}),
\\
\mathcal{N}_y &=& 
\sqrt{3} \left[
  d(\sigma_{XX} - \sigma_{YY})
  + e(\sigma_{YZ} - \sigma_{ZX})
\right],
\\
\mathcal{M}_x &=&
  b(2 \sigma_{ZZ} - \sigma_{XX} - \sigma_{YY}) \nonumber \\
  &+&c(2\sigma_{XY} - \sigma_{YZ} - \sigma_{ZX}),
\\
\mathcal{M}_y &=& 
\sqrt{3} \left[
  b(\sigma_{XX} - \sigma_{YY})
  + c(\sigma_{YZ} - \sigma_{ZX})
\right].
\eean
\end{subequations}
The relations between the hybrid-representation 
parameters
($a_1$, $a_2$, $b$, $c$, $d$, $e$)
and the NV-frame parameters
($g_{41}$, etc)
are given in the first two columns of Table \ref{tab:comparison}.
Importantly, $H_{\sigma0}$ and $H_{\sigma2}$ is identical
to the spin-stress Hamiltonian in Eqs. (1) and (2) of 
Ref.~\onlinecite{BarsonNL}.

In Table \ref{tab:comparison}, 
the DFT results for the cubic-frame spin-strain 
coupling-strength parameters are listed in the third column, 
whereas the experimental values\cite{BarsonNL} 
are listed in the fourth column. 
\footnote{
Note that with respect to the values quoted in 
Ref.~\onlinecite{BarsonNL}, the values in the fourth 
column of Table \ref{tab:comparison} have an inverted sign, 
because of the different sign convention for the stress
tensor: we assign a negative stress to compression.}
According to Table \ref{tab:comparison}, 
the signs of the DFT and experimental results are the same, 
and for all 4 parameters determined from the experiment, 
the order of magnitude matches well with that of the DFT result.
This suggests that the DFT method applied here captures
the key mechanism of interaction between the 
electron spin and the mechanical deformation, and 
gives confidence in the predictions for 
the previously omitted parameters $d$ and $e$.

\begin{table}
\caption{
Spin-stress coupling-strength parameters: Comparison of density functional theory and experimental~\cite{BarsonNL} results.
Parameters in the hybrid
representation ($a_1$, $a_2$, etc.)
are expressed in terms of the parameters in the NV-frame
representation ($g_{41}$, etc) in the second column. 
Par. and exp. are abbreviations for `parameters' and `experimental results'.}
\begin{ruledtabular}
\begin{tabular}{c|c|c|c}
par. & relation & DFT (MHz/GPa) & exp.\cite{BarsonNL} (MHz/GPa) \\\hline
$a_{1}$ & $\frac{2g_{41}+g_{43}}{3}$ & $-2.66\pm0.07$ & $-4.4\pm0.2$\\
$a_{2}$ & $\frac{-g_{41}+g_{43}}{3}$ & $2.51\pm0.06$ & $3.7\pm0.2$\\
$b$ & $\frac{-g_{15}+\sqrt{2}g_{16}}{12}$ & $1.94\pm0.02$ & $2.3\pm0.3$\\
$c$ & $\frac{-2g_{15}-\sqrt{2}g_{16}}{12}$ & $-2.83\pm0.03$ & $-3.5\pm0.3$\\
$d$ & $\frac{-g_{25}+\sqrt{2}g_{26}}{12}$ & $-0.12\pm0.01$ & -\\
$e$ & $\frac{-2g_{25}-\sqrt{2}g_{26}}{12}$ & $0.66\pm0.01$ & -
\end{tabular}
\end{ruledtabular}
\label{tab:comparison}
\end{table}

%=======

\section{Methods to measure
the spin-stress parameters}
\label{sec:exp}

To our knowledge, 
the spin-stress coupling strength parameters
$g_{25}$ and $g_{26}$ have not yet been measured.
In this section, we propose a method
that allows to determine those in an experiment which combines
the controlled application of mechanical stress and
ODMR.
The method, inspired by the experiment
of Ref.~\onlinecite{Broadway},
 requires a finite magnetic field 
along the NV axis, which tunes the system to the ground-state
level anticrossing (GSLAC)
where the $\ket{-1e}$ and $\ket{0e}$ electronic
states are approximately degenerate, 
$ B_z \approx B_g \equiv D/\gamma_e \approx 1024 \, \text{G}$.
In that setting, mechanical stress can induce
strong mixing of 
the spin eigenstates of the coupled electron-nuclear
system via the coupling-strength
parameters  $g_{25}$ and $g_{26}$.
In turn, the spin dynamics governed by this mixing
can be detected in a time-resolved fashion,
via photoluminescence-based optical
readout of the NV spin system.
First, in Section \ref{sec:effectivefield}, we introduce
our model, and show that the mechanical stress can be thought
of as an extra contribution to the external magnetic field, 
see Eq.~\eqref{effective_field}.
Second, in Section
\ref{sec:longitudinal}, we describe an 
arrangement that
can be used to determine the axial spin-stress coupling-strength
parameters $g_{41}$ and $g_{43}$.
Third, in Section \ref{sec:transverse}, we outline 
the experiment to determine the transverse coupling-strength
parameters $g_{25}$ and $g_{26}$.

\subsection{Effective magnetic field due to mechanical stress}
\label{sec:effectivefield}

\begin{figure}[b]
	\begin{flushleft}
		\includegraphics[width=\columnwidth]{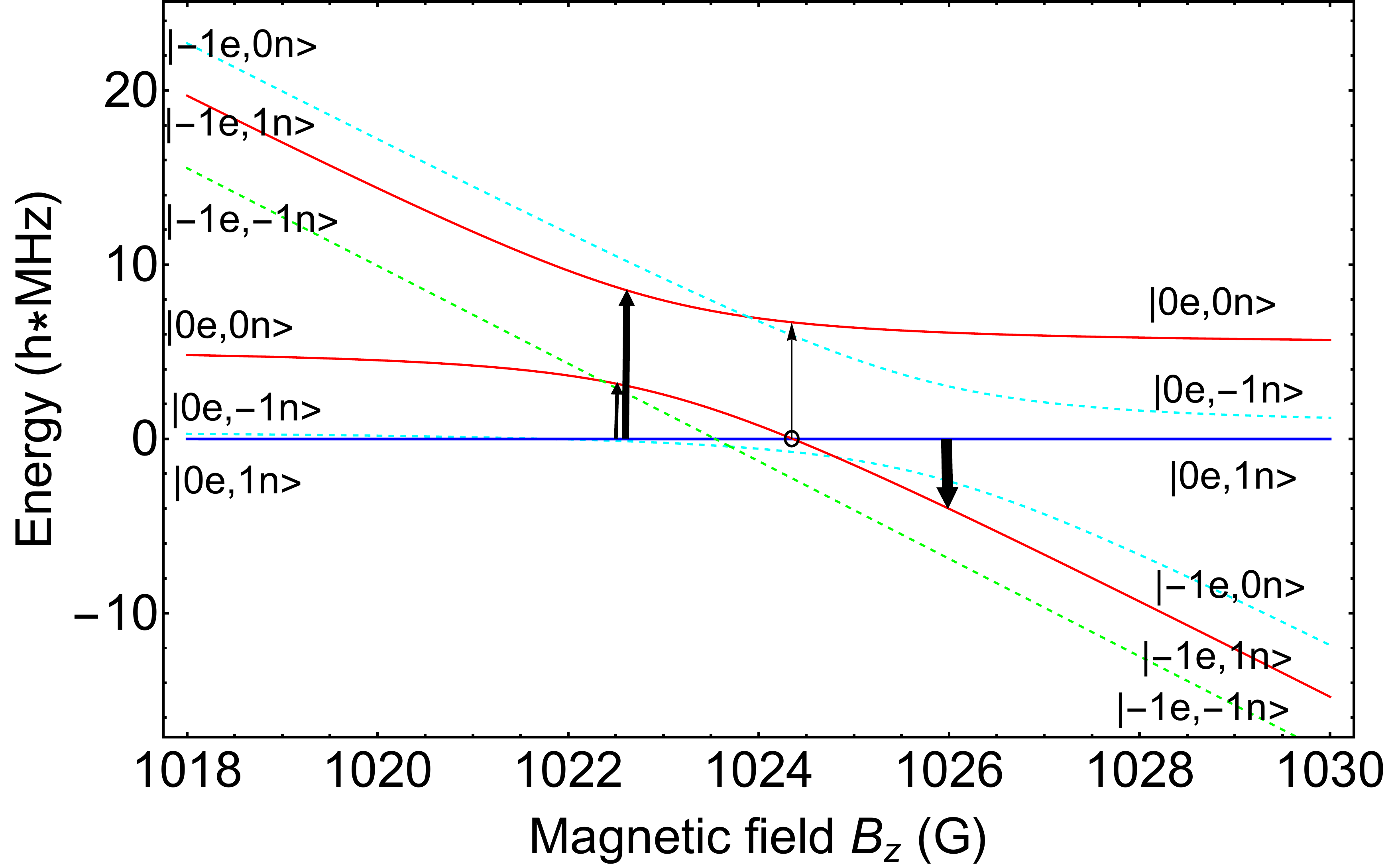}
		\caption{Level structure of the $^{14}$NV at the GSLAC as a function of the axial magnetic field $B_z$. 
		All transverse magnetic field components 
		and stress are zero, $B_x=B_y=0$, $\sigma=0$. 
		The circle marks the crossing that serves to identify the stress coupling coefficients $g_{25}$, $g_{26}$. 
		The levels coupled by the hyperfine interaction are shown with the same color (red solid; light blue dashed). 
		The arrows indicate the bright radiofrequency magnetic transitions at the corresponding values of the magnetic field. 
		In the absence of mechanical stress, the dashed 
		lines are invisible in optically detected magnetic resonance.
		}
		\label{Axial_Level_Structure}
	\end{flushleft}
\end{figure}

The measurement schemes described here work
in the vicinity of the GSLAC, where the 
$\ket{-1e}$ and $\ket{0e}$ electronic spin levels
are nearly degenerate.
This is where the stress-induced  terms of
$H_{\sigma 1}$,
which are
typically much smaller than the zero-field spin splitting $D$,
are most effective in mixing these two 
electronic spin states.  
Due to the presence of the N nuclear spin 
and hyperfine interaction, there is a hyperfine structure of
the energy spectrum at the GSLAC\cite{JacquesPRL2009}.
This is illustrated for the case of an $^{14}$N nuclear spin 
in 
Fig.~\ref{Axial_Level_Structure}:
instead of two electron spin levels crossing at $B_z = B_g$, 
there are six levels, with two level pairs showing 
hyperfine-induced anticrossings.
We focus on the case when the N atom of the NV 
center is an $^{14}$N; the analysis can be generalized
straightforwardly for the $^{15}$N case\cite{IvadyPRL2016,Broadway}.

%\subsubsection{Effective field description of NV center at GSLAC}

We assume that a magnetic field $B_z \approx B_g$ 
is applied, aligned with the NV axis.
Formally we write the magnetic field vector
as $\vec B = (B_x,B_y,B_z)$, but we will consider
only the case $B_x = B_y = 0$. 
The 9-dimensional Hamiltonian describing the 
coupled electron-nuclear system in the presence of 
the magnetic field and mechanical stress reads
\bean
H = H_e + H_\sigma + H_n + H_{hf},
\eean
where 
$H_e$ is defined in Eq.~\eqref{eq:he},
$H_\sigma$ is defined in Eq.~\eqref{Stress_ham},
$H_n$ describes 
the nuclear Zeeman effect
and the quadrupole moment of the $I=1$ spin of the
$^{14}$N via
\bean
H_n/h = - \gamma_n B_z I_z + Q I_z^2 ,
\eean
and $H_{hf}$ describes the hyperfine interaction via
\bean
H_{hf}/h=A_{||}S_zI_z+A_{\perp}(S_xI_x+S_yI_y).
\eean
We use the eigenstates of $I_z$ as the basis for the nuclear spin
states, labelled according to 
$I_z \ket{m_I n} = m_I \ket{m_I n} $, where
$m_I \in \{-1,0,+1\}$.
Note that in $H_n$ we use $\vec B = (0,0,B_g)$
for simplicity. 
The literature values of the coefficients~\cite{PhysRevB.79.075203} are $Q=-5.01$~MHz, $A_{||}=-2.14$~MHz, $A_{\perp}=-2.7$~MHz.

The six low-energy eigenstates of the 9x9 Hamiltonian 
$H$ are shown in Fig.~\ref{Axial_Level_Structure} as a function
of the axial magnetic field $B_z$, in the vicinity of the 
GSLAC.
For this plot, zero stress is assumed. 
Solid lines highlight the three levels that will be utilized to 
determine the spin-stress coupling-strength parameters. 
In Fig.~\ref{Axial_Level_Structure}, 
anticrossings are induced by hyperfine interaction, 
but far from the anticrossings the depicted
energy eigenstates are eigenstates of 
$S_z$ and $I_z$ to a good approximation, and therefore
are labelled accordingly, as $\ket{m_S e,m_I n}$.

When describing the effect of a nonzero mechanical 
stress, it is possible and helpful to introduce the notion of an 
\emph{effective magnetic field} vector
$(\Phi_x,\Phi_y,\Phi_z)$, 
which describes the
combined effect of the actual magnetic field and the 
stress-induced terms in the Hamiltonian.
To see this, let us first focus on the electronic degree of freedom
and the 2-dimensional low-energy electron spin subspace 
at the GSLAC. 
The electronic Hamiltonian in this 2-dimensional subspace
is expressed using the corresponding projector 
$P= |0e\rangle\langle 0e|+|\!-\! 1 e\rangle\langle -1 e|$
as
\begin{equation}
\begin{split}
P H_e P &= h \gamma_e\left(\begin{array}{cc}
0       & \frac{B_x-iB_y}{\sqrt{2}}  \\
 \frac{B_x+iB_y}{\sqrt{2}}        &B_g -B_z 
\end{array}\right).
\end{split}
\end{equation}
In the presence of a nonzero stress, described by the
matrix $\sigma$, this Hamiltonian generalizes to 
\begin{equation}
\begin{split}
P(H_{e}+H_{\sigma})P
&=h \gamma_e \left(\begin{array}{cc}
0       & \frac{\Phi_x-i\Phi_y}{\sqrt{2}}  \\
\frac{\Phi_x+i\Phi_y}{\sqrt{2}}        &B_g-\Phi_z 
\end{array}\right),
\end{split}
\end{equation}
where we introduced the effective magnetic field components
\begin{subequations}
\begin{eqnarray}
\Phi_x&=&B_x+\frac{g_{25}}{4\gamma_e}(\sigma_{xx}-\sigma_{yy})-\frac{g_{26}}{2\gamma_e}\sigma_{xz},
\label{eq:phix}
\\
\label{eq:phiy}
\Phi_y&=&B_y-\frac{g_{25}}{2\gamma_e}\sigma_{xy}-\frac{g_{26}}{2\gamma_e}\sigma_{yz},\\
\Phi_z &=&B_z-\frac{g_{41}}{\gamma_e}(\sigma_{xx}+\sigma_{yy})-\frac{g_{43}}{\gamma_e}\sigma_{zz}.
\label{eq:phiz}
\end{eqnarray}
\label{effective_field}
\end{subequations}
These expressions reveal that the mechanical 
stress can be thought of as an extra contribution to the applied
magnetic field. 

\subsection{Measuring the axial 
spin-stress parameters $g_{41}$, $g_{43}$}
\label{sec:longitudinal}

Our proposed experiment to determine $g_{41}$ and $g_{43}$ 
combines a controlled application of static uniaxial stress,
and optically detected magnetic resonance\cite{Broadway},
in an axial magnetic field that tunes the NV spin system
to the GSLAC.
Note that these coupling-strength parameters
have already been experimentally characterized 
by a different method 
in Ref.~\onlinecite{BarsonNL}.

The first stage of our proposed experiment is the observation
of certain parts of the hyperfine level structure 
shown in Fig.~\ref{Axial_Level_Structure}.
At this stage, no mechanical stress is applied. 
In the vicinity of the GSLAC, at
$(B_x,B_y,B_z) \approx (0,0,B_g)$,
the coupled electron-nuclear spin system is initialized 
to the state $\ket{\psi(0)} = \ket{0e,-1n}$ 
(blue solid line in Fig.~\ref{Axial_Level_Structure})
with an optical pulse.
Then, an ac magnetic pulse of a given frequency $f$, 
amplitude $B_\text{ac}$, and 
duration $\tau$ is applied.
On the one hand, 
if that magnetic pulse is off-resonant with respect to all
energy eigenstates in Fig.~\ref{Axial_Level_Structure}, then
the spin system remains in its initial state,
$\ket{\psi(\tau)} \propto \ket{0e,-1n}$.
Then, a readout optical pulse 
at time $t=\tau$ will result in significant photoluminescence
which is measured.
Note that the photoluminescence after the readout pulse 
is proportional to the occupation probability of the $\ket{0e}$ 
electron spin state, i.e., to the quantity
$\sum_{m_I } \left|\braket{0e,m_I n}{\psi(\tau)}\right|^2$.
On the other hand, if the magnetic field pulse is resonant 
with one of the 
transitions in Fig.~\ref{Axial_Level_Structure}, then it can
change the initial state to a state $\ket{\psi(\tau)}$ 
that contains a reduced weight
of the $\ket{0e}$ state, and thereby the photoluminescence
signal decreases.

To quantify this drop in the photoluminescence signal 
upon resonant excitation, we will use the 
quantity 
\bean
\label{eq:contrast}
C = 1- \sum_{m_I = -1,0,1} \left| \braket{0e, m_I n}{\psi(\tau)}  \right|^2,
\eean
and call it the \emph{photoluminescence contrast}. 
This quantity characterizes how effective the magnetic pulse
is in inducing spin transitions: the value of $C$ 
is zero for an off-resonant magnetic pulse, and can take
values between 0 and 1 for a resonant magnetic pulse.

\begin{figure}
	\includegraphics[width=5cm]{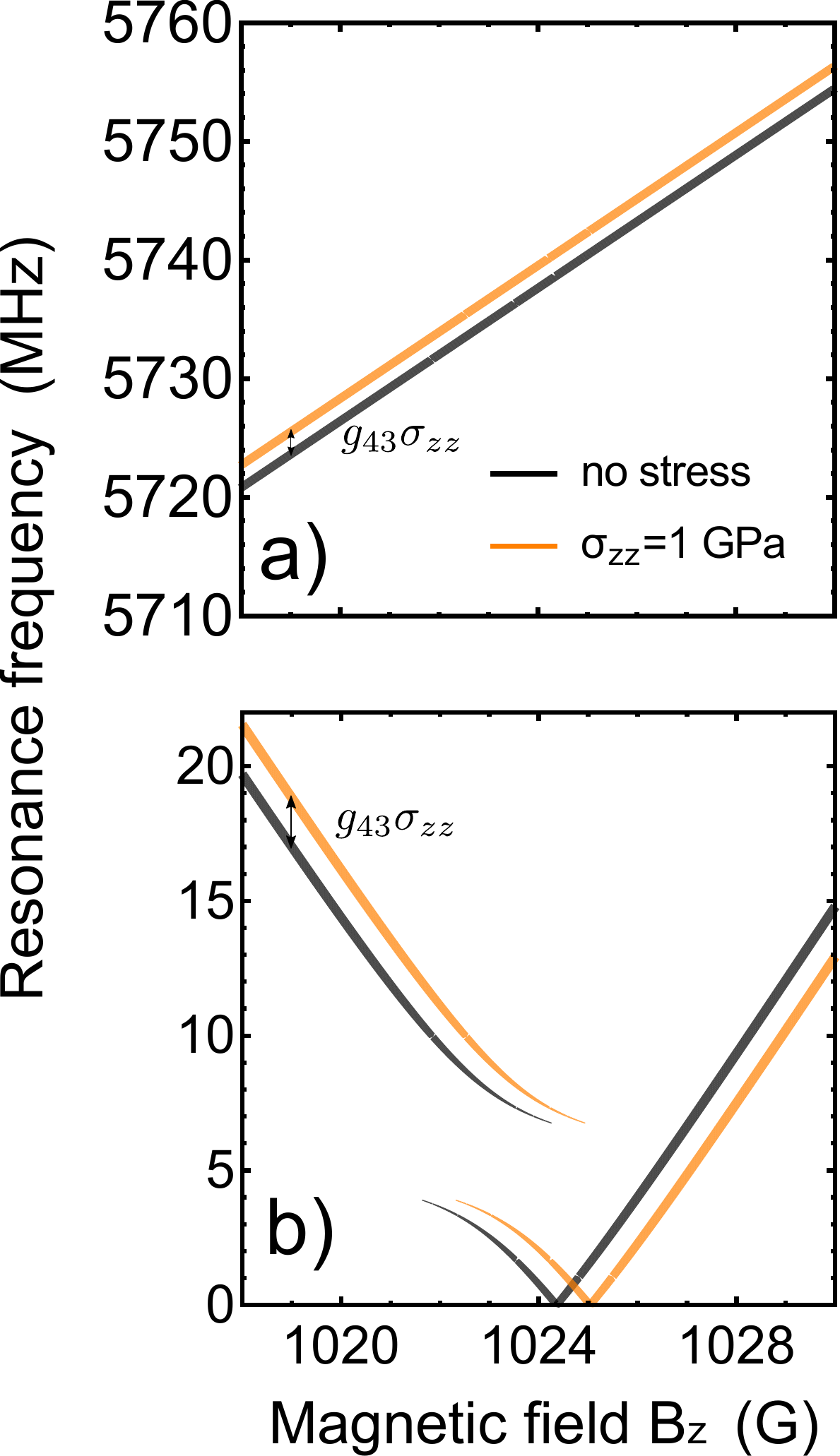}
	\caption{Effect of mechanical stress on the 
	photoluminescence contrast in optically detected magnetic 
	resonance of an $^{14}$NV centre.
	Black: no stress, orange: $\sigma_{zz} = 1\, \text{GPa}$.
	The curves show the dependence of hyperfine transition
	frequencies as function of the axial magnetic field $B_z$ 
	in the vicinity of the GSLAC.
	The thickness of each curve is proportional to the 
	photoluminescence contrast $C$ (Eq.~\ref{eq:contrast});
	maximal thickness corresponds to $C=1$.
	(a) High-energy transitions to $\ket{1e}$ spin states.
	(b) Low-energy transitions within the subspace
	of $\ket{0e}$ and $\ket{-1e}$. 
}
	\label{different stresses}
\end{figure}

The black curves in Fig.~\ref{different stresses}b visualize 
the predicted outcome of this experiment using the
photoluminescence contrast $C$, 
cf.~Fig.~2 of Ref.~\onlinecite{Broadway}.
Our Fig.~\ref{different stresses}b demonstrates that key features
of the hyperfine structure of the spin levels
of Fig.~\ref{Axial_Level_Structure} can be mapped using this
experimental technique.
To generate this plot, we calculated the five resonant 
transition frequencies from the spectral gaps in 
Fig.~\ref{Axial_Level_Structure}.
We plot these five curves in Fig.~\ref{different stresses}b, 
where the thickness of each curve is rescaled by 
the corresponding photoluminescence
contrast $C$. 
Hence, the black curves in Fig.~\ref{different stresses}b reveal that 
for a given magnetic field,
at most two out of the five transitions are bright.
The bright transitions at three specific $B_z$ values 
are also indicated in Fig.~\ref{Axial_Level_Structure}.
We calculated the photoluminescence contrast $C$ 
based on standard two-level 
Rabi dynamics in the rotating wave approximation, 
assuming resonant driving frequency $f$,
a magnetic pulse strength 
$b = g \mu_B B_\text{ac} \tau/h = \sqrt{2}/4$, 
and the ac magnetic field vector being aligned with the
$x$ axis. 
Note that the above pulse strength $b$ corresponds to 
an exact electron-spin $\pi$-pulse away from the GSLAC.

The second stage of the experiment is to repeat this 
ODMR spectroscopy in the presence of uniaxial 
z-directional strain, $\sigma_{zz} \neq 0$.
The predicted photoluminescence contrast for 
the case of $\sigma_{zz} = 1\, \text{GPa}$ 
is shown 
by the orange curves in Fig.~\ref{different stresses}b.
Apparently, the spectrum shifts along the $B_z$ axis. 
Measuring this shift reveals the spin-stress coupling-strength
parameter $g_{43}$.
In fact, simple analytical expressions can be obtained for
the locations of the ODMR resonances, including the
effect of the considered uniaxial strain. 
By projecting the 9x9 Hamiltonian $H$ to the
two-dimensional subspace of 
$\ket{0e,0n}$ and $\ket{-1e,1n}$,
and diagonalizing the resulting 2x2 Hamiltonian, 
we obtain the resonance frequencies corresponding to the
bright low-frequency transitions 
(blue $\to$ red transitions in Fig.~\ref{Axial_Level_Structure}):
\begin{eqnarray}
	\begin{split}
		&f_{\pm}=
		\left| \frac{1}{2}\left(-A_{||}+D\left(1+\frac{\gamma_n}{\gamma_e}\right)-Q-\gamma_e\Phi_z\right) \right.
             \\&
             \left. \pm\sqrt{A_{\perp}^2+\frac{1}{4}\left(A_{||}-D\left(1-\frac{\gamma_n}{\gamma_e}\right)-Q+\gamma_e\Phi_z\right)^2}
             \right| ,\\
	\end{split}
	\label{ODMR_radio_frequences}
\end{eqnarray}
For magnetic fields
significantly below the GSLAC,
e.g., around $B_z = 1019$ G 
in Fig.~\ref{different stresses}b,
the bright transition resonance frequency can 
be approximated by making a zeroth-order expansion
of $f_+$ 
(see Eq.~\eqref{ODMR_radio_frequences})
in $A_\perp$, and 
substituting Eq.~\eqref{eq:phiz} to the result, yielding
\bean
f_+ \approx 
-A_\parallel + D - \gamma_e B_z +g_{43} \sigma_{zz}.
\eean
This implies that $g_{43}$ can be directly calculated from
the measured 
stress-induced shift of the resonance frequency at a given
magnetic field (e.g., $B_z = 1019 \, \text{G}$) via
\bean
\label{eq:shiftz}
g_{43} = \frac{f_+(B_z,\sigma_{zz}) - f_+(B_z,0)}{\sigma_{zz}}.
\eean

The third, last stage of the experiment is to obtain
$g_{41}$ by
repeating this ODMR spectroscopy in the presence 
of uniaxial stress along $\vec n = (1,1,0)/\sqrt{2}$. 
In that case, the stress tensor reads 
$\sigma_{ij} = n_i n_j \sigma$, hence the three 
components 
$\sigma_{xx} = \sigma_{yy} = \sigma_{xy} = \sigma/2$ are
nonzero.
Because of the nonzero off-diagonal component
$\sigma_{xy}$, a nonzero effective magnetic field component
$\Phi_y$ is present, see Eq.~\eqref{eq:phiy},
seemingly complicating the previous analysis. 
However, assuming that our DFT predictions
in Table \ref{tab:DFTresults}
for the coupling-strength orders of magnitude are correct, 
the effect of this $\sigma_{xy}$-induced effective
transverse magnetic field component on the 
energy spectrum 
can be neglected away
from the anticrossing, e.g., at $B_z = 1019$ G.
Therefore, in this situation the stress-induced shift
of the resonance frequency 
can be translated to the coupling-strength parameter $g_{41}$ 
via 
\bean
\label{eq:shiftxy}
g_{41} = \frac{f_+(B_z,\sigma)-f_+(B_z,0)}{\sigma}.
\eean

We note that these coupling-strength coefficients
$g_{41}$ and $g_{43}$ can also be determined 
by utilizing the high-energy $\ket{1e,1n}$ spin state 
at the GSLAC and the corresponding 
$\sim$ 6 GHz ac magnetic field pulses.
This is illustrated by Fig.~\ref{different stresses}a,
where the photoluminescence contrast corresponding
to the $\ket{0e,1n} \to \ket{1e,1n}$ transition
is shown in the absence (black) and presence (orange)
of z-directional mechanical stress.
The relation between the coupling-strength parameters
and the shift of the resonance frequency is the same 
as for the low-energy transitions, see
Eqs.~\eqref{eq:shiftz} and \eqref{eq:shiftxy}.

\subsection{Measuring the transverse spin-stress parameters $g_{25}$, $g_{26}$}
\label{sec:transverse}

Here we propose and quantitatively analyze a method
for measuring the transverse spin-stress
coupling-strength coefficients $g_{25}$, $g_{26}$.
Similarly to the method 
in the preceding subsection,
this method also works in the vicinity of the GSLAC. 
It is based on the experiment 
discussed and implemented in 
Section IV of Ref.~\onlinecite{Broadway}, where 
Larmor-precession spin dynamics
was used to precisely measure the magnetic-field component
perpendicular to the NV axis (see, e.g., their Fig. 3).
Here we focus on how to measure the coupling strengths
$g_{25}$, $g_{26}$ in the case when the magnetic field is aligned with 
the NV axis.
Our method relies on the observation of 
Larmor-precession spin dynamics, which is affected by 
stress via the spin-stress interaction described
by Eq.~\eqref{Stress_ham}. 
The role of the transverse magnetic field components 
$B_x$ and $B_y$ in 
the experiment of Ref.~\onlinecite{Broadway} is
played by the stress-induced transverse effective magnetic field
components
$\Phi_x$ and $\Phi_y$ in our setup. 

First, recall the experimental scheme of Ref.~\onlinecite{Broadway}
for the special case when $B_z$ is tuned to the
blue-red level crossing in Fig.~\ref{Axial_Level_Structure}, 
$B_z = B_c$, 
denoted by a circle.
The two states that meet at the crossing are, to a good approximation,
$\ket{0e,1n}$ and
$\ket{\chi} = 
\frac{1}{\sqrt{1+\alpha^2}}(\alpha\ket{0e,0n}+\ket{-1e,1n})$,
where 
$\alpha=\frac{\gamma_eA_{\perp}}{Q\gamma_e-D\gamma_n}
\approx 0.5$.
For the readout, it will prove important that 
the weight of $\ket{\chi}$ in the $\ket{0e}$ 
subspace is 
$|\braket{0e,0n}{\chi}|^2\approx 0.2$, significantly lower than 1.
In the presence of a small transverse magnetic field, 
the blue-red level crossing in Fig.~\ref{Axial_Level_Structure} 
is split to an anticrossing, due to 
a coupling Hamiltonian matrix element 
between these states, which 
enters the two-level Hamiltonian of
$\ket{0e,1n}$ and $\ket{\chi}$ as
\bean
\label{eq:hlarmor}
H_L = \frac{h \gamma_e}{\sqrt{2(1+\alpha^2)}}
\left(\bna{cc} 
0 & B_x- i B_y \\
B_x + i B_y & 0
\eda \right).
\eean

In this setup, the 
experiment starts with an optical pulse that initializes
the spin system in $\ket{\psi(0)} = \ket{0e,1n}$ at $t=0$. 
Because of the finite transverse magnetic field in 
Eq.~\eqref{eq:hlarmor}, 
this initial state is not an energy eigenstate, and therefore the
time evolution $\ket{\psi(t)}$ exhibits 
complete Larmor-precession cycles
between the two states $\ket{0e,1n}$ and $\ket{\chi}$. 
To observe this Larmor precession,
the photoluminescence contrast $C(\tau)$ 
was measured\cite{Broadway} 
as a function 
of the waiting time $\tau$ following the initialization. 
This photoluminescence contrast 
$C(\tau)$ reveals the Larmor precession,
since the state $\ket{\chi}$ is mostly outside the $\ket{0e}$ 
subspace. 
The frequency of this Larmor
precession is derived from Eq.~\eqref{eq:hlarmor}:
\bean
\label{eq:broadwaylarmor}
f_L
= 
\sqrt{\frac{2}{1+\alpha^2}}
\gamma_e \sqrt{B_x^2+B_y^2} .
\eean

Here, we suggest to adopt this scheme to characterize
the effective transverse magnetic field components 
$\Phi_x$ and $\Phi_y$ defined
in Eq.~\eqref{effective_field}, and thereby measure 
the spin-stress coupling-strength coefficients $g_{25}$ 
and $g_{26}$. 
For simplicity, we make the following specifications. 
First, we take $B_x = B_y = 0$. 
Second, for an arbitrary uniaxial stress $\sigma_{ij} = \sigma n_i n_j$,
defined by its
direction $\vec n = (n_x,n_y,n_z)$ and magnitude $\sigma$,
we suggest to tune $B_z$ to the `virtual crossing point', 
i.e., to a value $B_z=  \tilde{B}_c$,  where the energy eigenvalues 
of $\ket{0e,1n}$ and $\ket{\chi}$ would be degenerate
in the virtual situation when the transverse
effective magnetic field components are turned off,
$\Phi_x = \Phi_y = 0$. 
That is guaranteed for
$\Phi_z = B_c$, which, together with Eq.~\eqref{eq:phiz}
implies 
\bean
 \tilde{B}_c = B_c + 
\frac{
g_{41}(n_x^2 +n_y^2) + g_{43} n_z^2}{\gamma_e}
\sigma.
\eean
This simple expression reveals that
this virtual crossing point can be identified once
the parameters $g_{41}$ and $g_{43}$ have been measured,
e.g., using the method of the preceding section. 

At this virtual crossing point, the role of the 
transverse effective magnetic fields $\Phi_x$ and
$\Phi_y$ is completely analogous to the role 
of $B_x$ and $B_y$ in Ref.~\onlinecite{Broadway}.
Namely, they force  the spin system initialized
in $\ket{0e,1n}$ to exhibit complete
Larmor precessions between the 
states $\ket{0e,1n}$ and
$\ket{\chi}$, with the Larmor frequency 
(cf. Eq.~\eqref{eq:broadwaylarmor})
\bean
f_L  =
\sqrt{\frac{2}{1+\alpha^2}} \gamma_e
\sqrt{\Phi_x^2+\Phi_y^2}.
\eean
From this, and using Eq.~\eqref{effective_field} for the effective magnetic fields, we find
\begin{equation}
\begin{split}
f_L =\sqrt{\frac{\frac{g_{25}^2}{4}n_{\perp}^4+g_{26}^2n_z^2n_{\perp}^2+g_{25}g_{26}n_xn_z(3n_y^2-n_x^2)}
{2(1+\alpha^2)}} |\sigma|,
\label{Larmor_frequency}
\end{split}
\end{equation} 
with $n_{\perp}=\sqrt{n_x^2+n_y^2}$.

Our result \eqref{Larmor_frequency} allows the identification of the
coefficients $g_{25}$ and $g_{26}$ by applying the uniaxial 
stress in different
directions and then measuring the Larmor precession frequency. 
For
example, 
the absolute value of
$g_{25}$ can independently be measured by applying the
uniaxial stress in the direction $\vec n =(1,1,0)/\sqrt{2}$. 
In that case,
Eq.~\eqref{Larmor_frequency} implies that 
this coupling-strength parameter is deduced from 
the measured Larmor frequency via
\bean
|g_{25}| = \sqrt{8(1+\alpha^2)} 
\frac{f_L}{|\sigma|} \approx 
3.17 \frac{f_L}{|\sigma|}.
\eean
Analogously, the absolute value of 
$g_{26}$ can independently be measured with the
uniaxial stress applied in the
direction $\vec n = (\sqrt{3},1,1)/\sqrt{5}$; 
for that case, we find
\bean
|g_{26}| = \frac{\sqrt{10(1+\alpha^2)}}{2} \frac{f_L}{|\sigma|}
\approx 
1.77 \frac{f_L}{|\sigma|}
\eean

We note that this procedure only allows us to determine the
absolute values of the
coupling-strength coefficients.
Nevertheless, it is straigthforward to generalize the above procedure
to determine the signs of the coefficients
by utilizing a finite tranverse magnetic field.
E.g., following up on our first example above, 
let us assume that we apply compressive uniaxial strain
$\sigma <0$
along $\vec n = (1,1,0)/\sqrt{2}$.
If the sign of $g_{25}$ is indeed negative, as
indicated by our DFT results in Table \ref{tab:DFTresults},
then the transverse effective magnetic field components 
read $\Phi_x = 0$ and $\Phi_y = B_y - p |\sigma|$, 
with $p>0$. 
Hence, according to Eq.~\eqref{Larmor_frequency},
the Larmor precession is slowed down
gradually as a magnetic field component along the y 
axis is switched on. 
On the other hand, if the sign of $g_{25}$ is positive, 
then a small y-directional magnetic field will speed up 
the Larmor precession.

\section{Discussion}
\label{sec:discussion}

\subsection{Potential applications}

\emph{Time-dependent mechanical deformation for
resonant spin control.}
Coherent spin control in NVs via ac
mechanical deformation has been demonstrated
with $\sim 1$ MHz Rabi frequency
for the magnetically forbidden
$\ket{-1e} \leftrightarrow \ket{1e}$ transition\cite{Barfuss:2015}. 
Our results imply that the other two, magnetically allowed,
transitions, $\ket{0e} \leftrightarrow \ket{\pm 1e}$, 
can also be induced in a similar fashion. 
This suggests that, in principle, the ac magnetic field used
routinely for spin control in NV-based experiments can be 
substituted by ac mechanical driving.
From the spin-strain Hamiltonian
$H_\varepsilon$ of Eq.~\eqref{eq:spinstrain},
we estimate that an ac strain $\varepsilon_{xx}$
with an amplitude of 0.01 
can provide mechanically induced Rabi oscillations
for the magnetically allowed transitions
with a Rabi frequency of $\sim 5 \, \text{MHz}$.

\emph{Time-dependent electric fields for resonant spin control.}
According to Eq.~\eqref{eq:electrichamiltonian}, 
an externally induced electric field interacts with the NV spin,
allowing for coherent electric control
of all three spin transitions of the NV. 
Electric control of the magnetically forbidden 
transition has been demonstrated in SiC\cite{Klimov}, 
but that of the magnetically allowed transitions has yet to be 
achieved. 
In Ref.~\onlinecite{Klimov}, electrical Rabi frequencies of $\sim 1$ MHz
were realized for the magnetically forbidden transition.
This Rabi frequency is proportional to the 
coupling-strength parameter $d_\perp$.
Furthermore, from
the dielectric strength of SiC it was estimated that 
$\sim 60$ MHz electrical Rabi frequencies should 
be reachable, comparable to magnetic spin control with
millitesla driving strength\cite{Jelezko_rabi,Fuchs_strongdriving}.
Noting that the $d_\perp$ parameter and
the dielectric strength are similar for NV centers in diamond,
and the $d'_\perp$ parameter is expected\cite{DohertyPRB2012} 
to be of the 
same order of magnitude as $d_\perp$, 
we speculate that 
the electrical Rabi frequencies for
the magnetically allowed transitions in diamond NVs centers 
could also reach a few tens of MHz.

\emph{Electrically driven, mechanically assisted
spin resonance using piezoelectric elements.}
Our results regarding the spin-strain coupling 
in $C_{3v}$ symmetric defects
promote a new way of using electric signals for coherent control,
for all three transitions between the spin-1 basis states.
Dynamical mechanical deformation can be created by 
ac electric fields (voltages) via piezoelectric elements
attached to the diamond crystal, e.g., a ZnO layer.
The functionality of such arrangements has already
been experimentally demonstrated using interdigital 
transducers serving as transmitters and receivers
of surface acoustic waves of the diamond 
crystal\cite{Golter_PRL2016, GolterPRX}.
The magnitude of strain created by the ac electric field
could further be enhanced using mechanical cavity resonators\cite{SchuetzPRX2015} for the surface 
acoustic waves.
The mechanical waves, when tuned to resonance with
the defect spin transition frequency, can then drive 
coherent spin Rabi oscillations. 
This working principle  allows for devices where
coherent control of the defect spins is performed 
via electrically driven, mechanically assisted spin resonance.

\subsection{Open problems}

\emph{Experimental characterization of the spin-strain 
and the spin-electric parameters.}
As discussed above, the spin-strain 
(spin-stress) coupling-strength parameters
of $H_{\varepsilon1}$ ($H_{\sigma1}$),
namely $h_{25}$ and $h_{26}$
($g_{25}$ and $g_{26}$, or $d$ and $e$, depending on the
representation), 
are yet to be characterized experimentally. 
Simliarly, the corresponding spin-electric
coupling-strength coefficient\cite{DohertyPRB2012} 
$d'_\perp$ in Eq.~\eqref{eq:electrichamiltonian}
is yet to be measured. 
We emphasize the technological relevance of these parameters:
the terms they multiply in the Hamiltonian can induce
magnetically allowed spin transitions, i.e., of the
$\ket{0e} \leftrightarrow \ket{\pm1 e}$ type;
therefore, for systems where these parameters are sufficiently strong,
 ac electric or ac mechanical driving could substitute the 
 ac magnetic field that is routinely used in most coherent spin-control
 experiments. 

\emph{Quantitative description of mechanically 
and electrically
driven electron spin resonance.}
The static spin-strain Hamiltonian \eqref{eq:spinstrain}
and the DFT-based
coupling-strength parameters in Table \ref{tab:DFTresults}
can be used to estimate the time scale (Rabi time) 
of spin control for an ac mechanical drive with a given
strain pattern. 
However, it is known from the theory of spin-orbit-mediated
electrically driven spin resonance\cite{Golovach,Crippa}, 
that even if an electric field does not modify 
the spin Zeeman splitting, 
it can induce transition between spin states.  
Hence it is expected that an accurate description 
of mechanically or electrically driven spin resonance 
for the NV, which probably involves
electronic spin-spin and spin-orbit interactions, 
requires a careful treatment of dynamical effects. 

\emph{Interaction of strain and electric fields with 
nuclear spins.}
The coherence time of the nuclear spin of the N atom in the NV
 exceeds that of the ground-state electronic spin,
and can be used as a long-lived quantum memory\cite{Fuchs:2011}.
Furthermore, the NV  can interact with $^{13}$C 
nuclear spins located in its vicinity. 
These highly coherent nuclear spins are heavily exploited in NV-based 
quantum-control experiments\cite{Childress,
Robledo:2011,Waldherr,Taminiau,Cramer,Broadway},
which is a strong motivation to understand the interaction 
of solid-state nuclear spins with electric and strain fields.
Important steps in this direction have already been taken\cite{Ogura,Thiele,Sigillito,Tosi-nuclearspin,Boross-nuclearspin,Mansir}, 
but the experimental and theoretical 
characterization of the spin-electric and spin-strain 
interactions for NV  nuclear spins
is yet to be done.

We anticipate that the nature of the problem is qualitatively different for
(i) a spin-1/2 nuclear spin,
e.g., of a $^{15}$N or a $^{13}$C atom,  
and 
(ii) a nuclear spin that is
larger than 1/2, e.g., of a $^{14}$N atom.
In case (i) the nuclear spin does not interact 
directly with electric or strain fields\cite{Slichter}.
However, these fields do interact with the 
electronic spin, which can serve as a quantum transducer
that translates these fields to the nucleus via the hyperfine
interaction\cite{Childress,Sangtawesin,Chen,Thiele,Sigillito,Boross-nuclearspin} (Knight field).
In case (ii), the nuclear spin has a nonzero 
electric quadrupole moment, and therefore can interact
directly with electric and strain fields via the local electric-field
gradient\cite{Slichter,Sigillito}.
Then, the 
direct interaction and the hyperfine-mediated interaction
will compete. 
In both cases (i) and (ii), the results of our present work
can serve as a starting point to evaluate the 
hyperfine-mediated contribution.

\section{Conclusions}

We have established the spin-strain and spin-stress interaction
Hamiltonians for the NV ground state, and
numerically determined the six independent parameters of this
Hamiltonian using density functional theory.
Focusing on the new Hamiltonian term $H_{\varepsilon1}$ 
identified in this work, 
we proposed an NV-based 
experimental setup where spin effects caused by a 
static mechanical deformation can be observed,
and suggested coherent mechanical or electric spin control of the 
the magnetically allowed spin transitions. 
All qualitative considerations of this work should hold 
for the whole family of defects with $C_{3v}$ symmetry 
and spin-1 electronic states.

\acknowledgments
We thank A. Auer,  
M. Barson, M. Doherty, A. Falk, J. Heremans, V. Iv\'ady, J. Michl, 
S. Sangtawesin, D. Szaller, 
G. Thiering, and B. Zhou
for helpful discussions.
AP is supported by the National Research Development and Innovation Office of Hungary (NKFIH) Grants
105149
%zarand
and 
%asboth
124723,
and the \'UNKP-17-4-III 
New National Excellence Program of the Ministry of 
Human Capacities of Hungary. AG thanks for the support of the EU
Commission in the DIADEMS project (Grant No.~611143) 
and NKFIH within the Quantum Technology National Excellence Program  (Project No.~2017-1.2.1-NKP-2017-00001).
VOS and GB are supported by the DFG within the collaborative
research center SFB 767.

\appendix

%===

\section{Symmetry analysis of the spin-strain Hamiltonian}
\label{app:symmetry}

In this Appendix, we describe two derivations of the symmetry-allowed
spin-strain Hamiltonian $H_\varepsilon$ of Eq.~\eqref{eq:spinstrain}.
The first derivation is an elementary one, without reference to group representation theory,
whereas the second one builds upon concepts of the latter.
The two methods yield the same result Eq.~\eqref{eq:spinstrain}.

\subsection{Elementary derivation}
Our goal is to find the most general form of the 
Hamiltonian describing the interaction between
a homogeneous strain and the ground-state 
spin (spin-1) of the NV.
More precisely, we aim at finding the most general form
of the interaction that is
(i) allowed by the requirement of time reversal symmetry,
(ii)  allowed by the spatial symmetries 
($C_{3v}$) of the structure,
(iii) linear in the elements of the strain tensor $\varepsilon$.

The interaction Hamiltonian should be quadratic in the 
components of the spin vector
$\vec S = (S_x,S_y,S_z)$, 
as time reversal symmetry changes the
sign of those, and the interaction Hamiltonian should be
invariant upon time reversal.
Our $\vec S$ is dimensionless, fulfilling
$S^2 = 2$.

Therefore, our 
starting point is the Hamiltonian 
\bean
H_\varepsilon = \sum_{\alpha,\beta,\gamma,\delta \in \{x,y,z\}}
h_{\alpha\beta\gamma\delta} S_\alpha S_\beta 
\varepsilon_{\gamma\delta},
\eean
where $h$ is a four-dimensional matrix
with real entries. 
Apparently, $h$ has 81 independent elements;
this will now be reduced, first without invoking
any symmetries of the considered system. 

To this end, we 
exploit the fact the 9-element set
$\{S_\alpha S_\beta | \alpha, \beta \in \{x,y,z\}\}$
is overcomplete (linearly dependent) 
in the six-dimensional vector space of 
$3\times 3$ Hermitian time-reversal invariant matrices. 
A six-element basis of that vector space is provided by, e.g., 
$( 1,
\frac 1 2 \{S_x ,S_y\},
\frac 1 2 \{S_y ,S_z\},
\frac 1 2 \{S_z ,S_x\},
S_z^2,
S_x^2-S_y^2
)
\equiv (\Sigma_0, \Sigma_1, \dots, \Sigma_5)$.
We will neglect the unit matrix $\Sigma_0$ from now.
Furthermore, we will refer to $\Sigma$ as a map
$(S_x,S_y,S_z) \mapsto \Sigma(\vec S) :=
(\Sigma_1,\Sigma_2, \dots, \Sigma_5)$.

A further simplification is allowed by the fact that 
the strain tensor is symmetric.
Therefore 
it can be thought of as a 
six-dimensional column vector, 
$\epsilon = (\varepsilon_{xx},\varepsilon_{yy},\varepsilon_{zz},
\varepsilon_{yz},\varepsilon_{zx},\varepsilon_{xy})^T$.
We will consider $\epsilon$ as a function
that maps the strain tensor to a six-dimensional vector, 
$\varepsilon \mapsto \epsilon(\varepsilon)$.

Using these simplifications, 
we can express the most
general Hamiltonian as 
\bean
H_\varepsilon = 
\sum_{n=1}^5 \sum_{v=1}^6
h_{nv} \Sigma_n \epsilon_v,
\eean
where $h$ is a $5\times 6$ matrix with real entries, 
i.e., it is characterized by only 30 independent elements. 

We will now further reduce this number using the
spatial symmetry of the NV.
Its symmetries are the isometries 
in the group $C_{3v}$. 
Those are generated by a 3-fold rotation around the 
$z$ axis, $\mathcal{R}$, and 
the reflection on the $xz$ plane, 
$\mathcal{M}$.
These isometries are represented on a position vector
by the $3\times 3$ matrices 
\bean
R = \left(\bna{ccc}
\cos \frac{2\pi}{3} & -\sin \frac{2\pi}{3} & 0 \\
\sin \frac{2\pi}{3} & \cos \frac{2\pi}{3} & 0 \\
0 & 0 & 1
\eda\right),
\eean
and
\bean
M = \left(\bna{ccc}
1 & 0 & 0 \\
0 & -1 & 0 \\
0 & 0 & 1
\eda\right),
\eean
respectively. 

A point isometry transforming the structure also transforms
the associated physical quantities. 
For us, one of the relevant quantities
is the strain tensor, which is transformed as
$\varepsilon \mapsto R \varepsilon R^{-1}$
and 
$\varepsilon \mapsto M \varepsilon M^{-1}$.
The other relevant quantity is the spin vector, which transforms
as a pseudovector (or axial vector).
That is, the rotation is represented on the spin
as $\vec S \mapsto R \vec S$, 
but the reflection is 
represented as $\vec S \mapsto M' \vec S$
with 
\bean
M' = \left(\bna{ccc}
-1 & 0 & 0 \\
0 & 1 & 0 \\
0 & 0 & -1
\eda \right).
\eean

We require that the Hamiltonian is invariant against the
transformations  of the point group of the structure;
formally that is written as 
\bean
\sum_{n=1}^5 \sum_{v=1}^6
h_{nv} \Sigma_n(\vec S) \epsilon_v( \varepsilon )
=
\sum_{n=1}^5 \sum_{v=1}^6
h_{nv} \Sigma_n(R \vec S) \epsilon_v(R \varepsilon R^{-1}),
\nonumber \\
\eean
and
\bean
\sum_{n=1}^5 \sum_{v=1}^6
h_{nv} \Sigma_n(\vec S) \epsilon_v( \varepsilon )
=
\sum_{n=1}^5 \sum_{v=1}^6
h_{nv} \Sigma_n(M' \vec S) \epsilon_v(M \varepsilon M^{-1}),
\nonumber
\\
\eean
Both of these equations form a homogeneous linear set
of 30 equations, with the 30 $h_{nv}$ coupling-strength coefficients
being the unknowns.
Hence these equations establish linear relationships
between the various $h_{nv}$ coefficients, that is, 
they reduce the number of free parameters
in the Hamiltonian.

These equations can be solved, 
e.g., symbolically using computer algebra. 
Inserting the solutions
to $H_\varepsilon$ yields our symmetry-allowed spin-strain
interaction Hamiltonian of
Eq.~\eqref{eq:spinstrain}.

\subsection{Derivation based on group representation theory}
The $C_{3v}$ symmetry group of the NV
has three irreducible representations (irreps):
the trivial 1D irrep $A_1$, the 1D irrep $A_2$ and 
the 2D irrep $E$. 
The quadratic spin-component combinations that transform according to the
trivial $A_1$ irrep are
\bean
f_{A_1,1}^{(\text{spin})}& =&
S_x^2 + S_y^2, \\
f_{A_1,2}^{(\text{spin})} &=&
S_z^2.
\eean
Analogously, the linear strain-component combinations tranforming as $A_1$ are
\bean
f_{A_1,1}^{(\text{strain})}& =&
\varepsilon_{xx} +\varepsilon_{yy}, \\
f_{A_1,2}^{(\text{strain})} &=&
\varepsilon_{zz}.
\eean
We will refer to the number of these combinations as $n(A_1) =2$.
There are no such combinations transforming according to $A_2$,
i.e., $n(A_2) = 0$.
The quadratic spin-component combinations forming 2D vectors,
which transform according to the 2D irrep $E$, are
\bean
f_{E,1}^{(\text{spin})} &=&
	\left(\bna{c}
		S_x^2 - S_y^2 \\
		-\{S_x,S_y\}
	\eda\right), \\
f_{E,2}^{(\text{spin})} &=&
	\left(\bna{c}
		\{S_x,S_z\} \\
		\{S_y,S_z\}
	\eda\right).
\eean
Analogously, the linear strain-component combinations forming 2D vectors,
which transform according to $E$, are
\bean
f_{E,1}^{(\text{strain})} &=&
	\left(\bna{c}
		\varepsilon_{xx}-\varepsilon_{yy} \\
		-2\varepsilon_{xy}
	\eda\right), \\
f_{E,2}^{(\text{strain})} &=&
	\left(\bna{c}
		\varepsilon_{xz} \\
		\varepsilon_{yz}
	\eda\right).
\eean
These imply $n(E) = 2$.

The symmetry-allowed spin-strain Hamiltonian 
is an arbitrary linear combination of the scalar products of the above-defined
(1D and 2D) vectors that transform according to the same irrep. 
Formally, this is written in a compact fashion as follows:
\bean
\label{eq:hfromirreps}
H_\varepsilon =
\sum_{\Gamma \in \text{irreps}}
\sum_{\sigma,\tau = 1}^{n(\Gamma)}
c_{\Gamma \sigma \tau} 
\left(
  f_{\Gamma \sigma}^{(\text{spin})} \cdot f_{\Gamma \tau}^{(\text{strain})}
\right).
\eean
Here, the quantites $c_{\Gamma \sigma \tau}$ are independent real coefficients
(coupling-strength parameters)
that are not constrained by symmetry, and can be determined from microscopic
models or experiments, as discussed in the main text. 
According to the counts of the previous paragraph, the
sum in Eq.~\eqref{eq:hfromirreps} has 8 terms, and therefore
there are 8 independent 
coupling-strength coefficients. 
However, since a uniform energy shift of the spin states 
in the Hamiltonian can be disregarded,
and $f_{A_1,1}^{(\text{spin})}$
and  $f_{A_1,1}^{(\text{spin})}$ do add up to a constant 
due to $S_x^2+S_y^2+S_z^2 = 2$,
we can set $c_{A_1,1,1} = c_{A_1,1,2} = 0$ without the loss
of generality.
This implies that there are six independent nonzero coupling-strength
parameters.

Direct evaluation of the terms in Eq.~\eqref{eq:hfromirreps}
and comparison with Eq.~\eqref{eq:spinstrain}
allows to establish the relations between
the coupling-strength coefficients:
\begin{subequations}
\bean
c_{A_1,2,1} &=& h_{41}, \\
c_{A_1,2,2} &=& h_{43}, \\
c_{E,1,1} &=& \frac 1 4  h_{15}, \\
c_{E,1,2} &=& - \frac 1 2 h_{16}, \\
c_{E,2,1} &=& - \frac 1 4 h_{25} ,\\
c_{E,2,2} &=& \frac 1 2 h_{26}.
\eean
\end{subequations}

%===
\section{Computing spin-strain parameters 
with density functional theory}
\label{app:dft}

We  determined the 
spin-strain coupling-strength parameters 
using numerical DFT calculations. 
We applied DFT for electronic structure calculation 
combined with geometry optimization,
using the PBE functional\cite{PBE} 
in the plane-wave-based 
Vienna Ab initio Simulation Package (VASP)\cite{VASP1,VASP2,VASP3,VASP4}. 
The core electrons were treated in the projector augmented-wave 
(PAW) formalism\cite{paw}. 
The calculations were performed with $600~\text{eV}$ 
plane wave cutoff energy. 
The model of the NV in bulk diamond was constructed 
using a 512-atom diamond simple cubic supercell within the 
$\Gamma$-point approximation. 
We use a negative sign convention for compressive strain.
To model the structure subject to mechanical strain, 
described by the strain tensor $\varepsilon$, 
we deform the cubic supercell to a parallelepiped, 
whose edge vectors are obtained by transforming the
undeformed edge vectors with the matrix 
$1+\varepsilon$ in the cubic reference frame,
and allow the atomic positions to relax.
For each strain configuration, 
the elements of the 
$3 \times 3$
zero-field splitting matrix $D$,
defining the ground-state spin Hamiltonian
via $H = \vec S^T \cdot  D \cdot \vec S$, 
were calculated using the VASP
implementation by Martijn Marsman with
the PAW formalism~\cite{Bodrog}.

\begin{figure}
\includegraphics[width=\columnwidth]{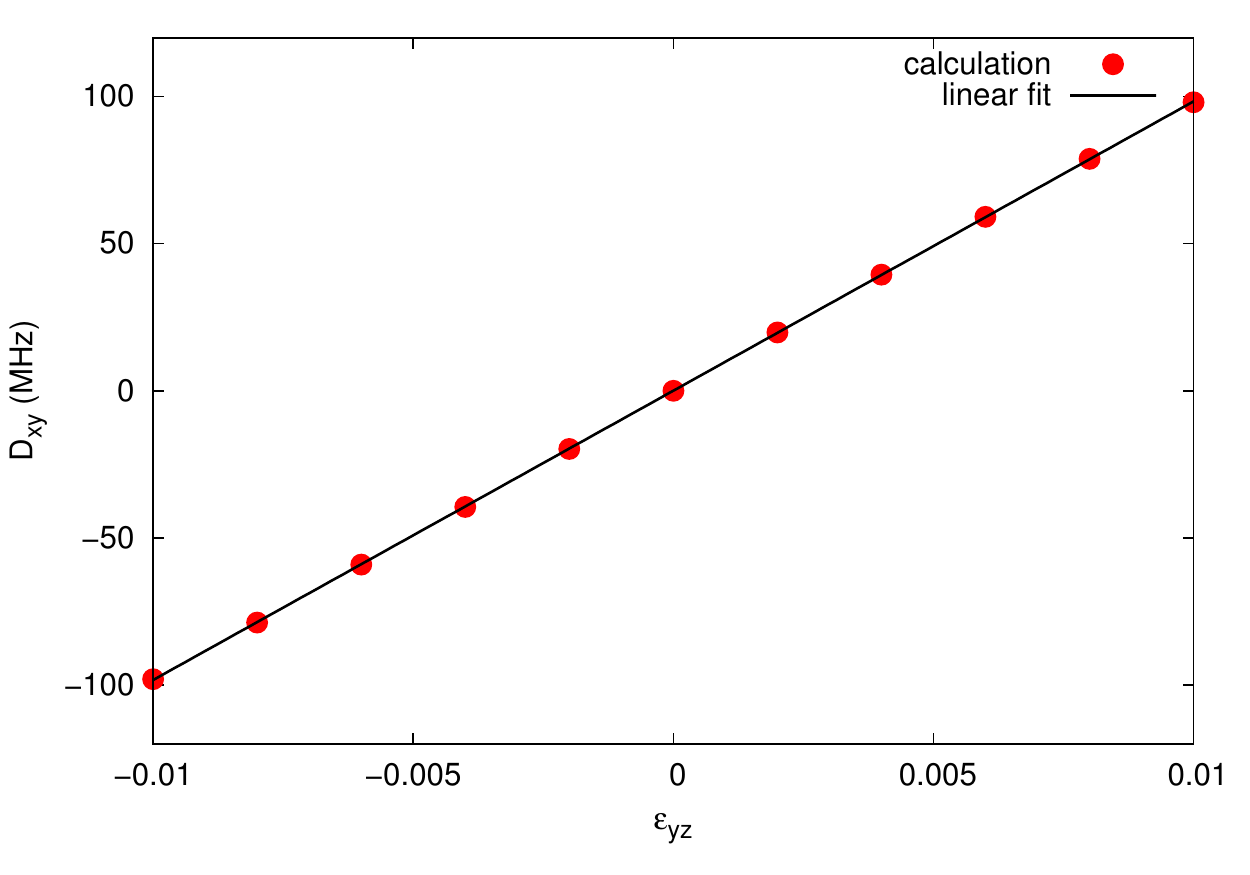}
\caption{\label{fig:h16_yz}Strain dependence of the zero-field splitting
matrix element $D_{xy}$. 
Data points show the DFT results for the
matrix element $D_{xy}$, as a
function of the strain component $\varepsilon_{xy}$, 
with all other strain components set to zero. 
Solid line shows a linear fit, with 
a slope of  $9832\pm9 ~\mathrm{MHz/strain}$,
allowing to obtain the coupling-strength parameter
$h_{16}$ via Eq.~\eqref{eq:dtoh}.}
\end{figure}

We illustrate our methodology to obtain the six spin-strain
coupling-strength coefficients with the example of $h_{16}$. 
To determine $h_{16}$, we deform
the supercell using a strain tensor whose only 
nonvanishing element is $\varepsilon_{yz}$,
and obtain the $D$ matrix from the calculation.
Due to Eq.~\eqref{eq:spinstrain}, 
the chosen strain configuration implies that
the Hamiltonian has the form
\bean
H = 
\frac 1 2 \varepsilon_{yz}
\vec{S}^T \cdot \left( \bna{ccc}
0 &  h_{16}  & 0 \\
h_{16} & 0 &  h_{26} \\
0 &  h_{26} & 0
\eda \right)
\vec S.
\eean
This, together with the above definition of the $D$ matrix, 
yields 
\bean
\label{eq:dtoh}
h_{16} = 2 
\left.
	\frac{\partial D_{xy}}{\partial \varepsilon_{yz}}
\right|_{\varepsilon = 0}.
\eean
To be able to estimate the numerical error of our DFT calculations,
we infer the derivative in Eq.~\eqref{eq:dtoh} using a sequence
of calculations with 11 equidistant values of 
$\varepsilon_{yz}$ between -0.01 and 0.01.
The resulting $D_{xy}(\varepsilon_{yz})$
data points are shown in Fig.~\ref{fig:h16_yz}.
From a linear fit, shown as the solid line in Fig.~\ref{fig:h16_yz}, 
we infer the coupling-strength
coefficient $h_{16}$ via Eq.~\eqref{eq:dtoh} and
its standard deviation.

Similar procedures can be applied to determine the remaining
five coupling-strength parameters, and the results
are shown in Table \ref{tab:DFTresults}, with the following remarks.
(i) To obtain the value of $h_{41}$ and its error in 
Table \ref{tab:DFTresults}, we calculated the corresponding
results from the $\varepsilon_{xx}$ dependence of the $D$ matrix,
as well as from its $\varepsilon_{yy}$ dependence, 
and averaged these results.
(ii) We used similar averaging in the case of $h_{26}$ and $h_{16}$,
which we calculated from 
from the $\varepsilon_{xz}$ dependence,
as well as from its $\varepsilon_{yz}$ dependence
(the latter is illustrated in Fig.~\ref{fig:h16_yz}).
(iii) 
We determined the values for $h_{25}$ and $h_{15}$ 
from the $\varepsilon_{xy}$ dependence.
In Table \ref{tab:DFTresults}, we also present the 
spin-stress parameters ($g_{41}$, etc),
 which we determined from the DFT-based spin-strain 
 parameters using the conversion procedure detailed in 
 Appendix \ref{app:conversion}.

%===

\section{Converting spin-strain parameters to 
spin-stress parameters}\label{app:conversion}

To calculate the spin-stress coupling-strength parameters
in Table \ref{tab:DFTresults}
from the DFT-based spin-strain parameters,
we start from the stiffness tensor $C$ of bulk diamond, 
and take the following values\cite{Kaxiras}
for its elements in the cubic reference frame:
$C_{11}=1076~\mathrm{GPa}$, $C_{12}=125~\mathrm{GPa}$, $C_{44}=576~\mathrm{GPa}$.
First, we transform the stiffness tensor to the NV frame;
we denote the resulting $6\times 6$ stiffness matrix 
in the Voigt notation as $C$. 
To convert our spin-strain Hamiltonian 
Eq.~\eqref{eq:spinstrain} 
to spin-stress Hamiltonian,
we express the strain components in Eq.~\eqref{eq:spinstrain}
using stress components via
$\varepsilon = C^{-1} \sigma$, 
where $\varepsilon = (\varepsilon_{xx},\varepsilon_{yy},\varepsilon_{zz},
2\varepsilon_{yz},2\varepsilon_{zx},2\varepsilon_{xy})$
and $\sigma = (\sigma_{xx},\sigma_{yy},\sigma_{zz},\sigma_{yz},\sigma_{zx},\sigma_{xy}) $ are now also in Voigt notation; 
note the factor of two in front of the off-diagonal strain
components.

The inverted stiffness tensor in the NV frame reads
\begin{equation}
C^{-1}=\left(\begin{matrix}
C^{-1}_{11} & C^{-1}_{12} & C^{-1}_{13} & 0 & C^{-1}_{15} & 0 \\
C^{-1}_{12} & C^{-1}_{11} & C^{-1}_{13} & 0 & -C^{-1}_{15} & 0 \\
C^{-1}_{13} & C^{-1}_{13} &C^{-1}_{33} & 0 & 0 & 0\\
0 & 0 & 0 & C^{-1}_{44} & 0 & C^{-1}_{46} \\
C^{-1}_{15} & -C^{-1}_{15} & 0 & 0 & C^{-1}_{44} & 0\\
0 & 0 & 0 & C^{-1}_{46} & 0 & C^{-1}_{66}\\
\end{matrix}\right),
\end{equation}
yielding the following following expressions for the
spin-stress parameters:
\begin{subequations}
\begin{align}\label{conversion_beg}
g_{41}&=h_{41}\left(C^{-1}_{11}+C^{-1}_{12}\right)+h_{43}C^{-1}_{13}
\\
g_{43}&=2h_{41}C^{-1}_{13}+h_{43}C^{-1}_{33}
\\
g_{26}&=h_{26}\frac{1}{2}C^{-1}_{44}-h_{25}C^{-1}_{15}
\\
g_{25}&=h_{25}\left(C^{-1}_{11}-C^{-1}_{12}\right)-h_{26}C^{-1}_{15}
\\
g_{16}&=h_{16}\frac{1}{2}C^{-1}_{44}-h_{15}C^{-1}_{15}
\\
g_{15}&=h_{15}\left(C^{-1}_{11}-C^{-1}_{12}\right)-h_{16}C^{-1}_{15}
\label{conversion_end}
\end{align}
\label{eq:conversion}
\end{subequations}
These relations, together with the numerical values 
 of the inverse stiffness matrix elements,
\begin{align*}
C^{-1}_{11}&=86\cdot10^{-5}~\mathrm{1/GPa}, &C^{-1}_{33}&=83\cdot10^{-5}~\mathrm{1/GPa}, \\ C^{-1}_{44}&=198\cdot10^{-5}~\mathrm{1/GPa}, &C^{-1}_{66}&=186\cdot10^{-5}~\mathrm{1/GPa},\\ C^{-1}_{12}&=-7\cdot10^{-5}~\mathrm{1/GPa}, &C^{-1}_{13}&=-4\cdot10^{-5}~\mathrm{1/GPa},\\ C^{-1}_{15}&=9\cdot10^{-5}~\mathrm{1/GPa}, &C^{-1}_{46}&=-17\cdot10^{-5}~\mathrm{1/GPa},
\end{align*}
are used to obtain the $g_{41}$, etc values in 
Table \ref{tab:DFTresults}.

\bibliography{nvstrain}

\end{document}